\documentclass[11pt,twoside]{article}
\usepackage{./asp2014}

\aspSuppressVolSlug
\resetcounters

\begin{document}

\title{The JScanam Map-Maker Method Applied to Herschel/PACS Photometer Observations}
\author{Javier~Graci{\'a}-Carpio,$^1$ Michael~Wetzstein,$^2$ H{\'e}l{\`e}ne~Roussel$^3$, and the~PACS Instrument Control Centre Team
\affil{$^1$Max-Planck-Institut f{\"u}r Extraterrestrische Physik, Garching, Munich, Germany; \email{jgracia@mpe.mpg.de}}
\affil{$^2$University Observatory Munich, Munich, Germany; \email{drwstein@usm.lmu.de}}
\affil{$^3$Institut d'Astrophysique de Paris, Paris, France; \email{roussel@iap.fr}}}

\paperauthor{Javier Graci{\'a}-Carpio}{jgracia@mpe.mpg.de}{}{Max-Planck-Institut f{\"u}r Extraterrestrische Physik}{IR/sub-mm group}{Garching}{Munich}{85748}{Germany}
\paperauthor{Michael Wetzstein}{drwstein@usm.lmu.de}{}{University Observatory Munich}{}{Munich}{Munich}{81679}{Germany}
\paperauthor{H{\'e}l{\`e}ne Roussel}{roussel@iap.fr}{}{Institut d'Astrophysique de Paris}{}{Paris}{Paris}{75014}{France}

\begin{abstract}
JScanam is the default map-maker for Herschel/PACS photometer observations. Making use of the redundant information from multiple passages on the sky with different scanning directions, JScanam is able to remove the $1/f$ noise that severely affects PACS far-infrared maps, preserving at the same time point sources and real extended emission. The JScanam pipeline has been designed to run automatically on all kind of maps and astronomical environments, from Galactic star-forming clouds to deep cosmological fields. The results from the JScanam automatic pipeline can be easily inspected and downloaded from the Herschel Science Archive and the new ESA Sky interface.
\end{abstract}

\section*{Herschel and the PACS photometer}
The Herschel satellite \citep{Pilbratt2010} was launched in May 2009 and during its almost 4 years of technical and scientific operations it observed up to 10\% of the far-infrared sky. The studied regions cover a wide range of astronomical environments, from Galactic star-forming clouds, to galaxies in the local and high-redshift Universe.
   
The Photodetector Array Camera and Spectrometer \citep[PACS,][]{Poglitsch2010} was one of the three science instruments onboard Herschel. In its photometric scanning mode, PACS was able to observe wide areas of the sky in three far-infrared bands centered at 70, 100 and 160$\mu$m. Among other important discoveries, PACS maps revealed the filamentary structure of the cold interstellar medium \citep{Andre2010} and resolved $\sim$80\% of the Cosmic Far-infrared Background \citep{Berta2010}. 

\section*{PACS signal properties}
Since the PACS photometer detectors are bolometers, their signal is dominated by the so-called $1/f$ noise. This noise has higher amplitude at longer timescales, and results in visible signal drifts that affect all spatial structures in the maps (see figures~\ref{f1_p043} and \ref{f2_p043}). High-pass filtering techniques can be used to remove this kind of noise, leaving point sources unaffected \citep{Popesso2012}. Unfortunately, these techniques also tend to remove a large fraction of the real astronomical extended emission from the maps.

\articlefigure{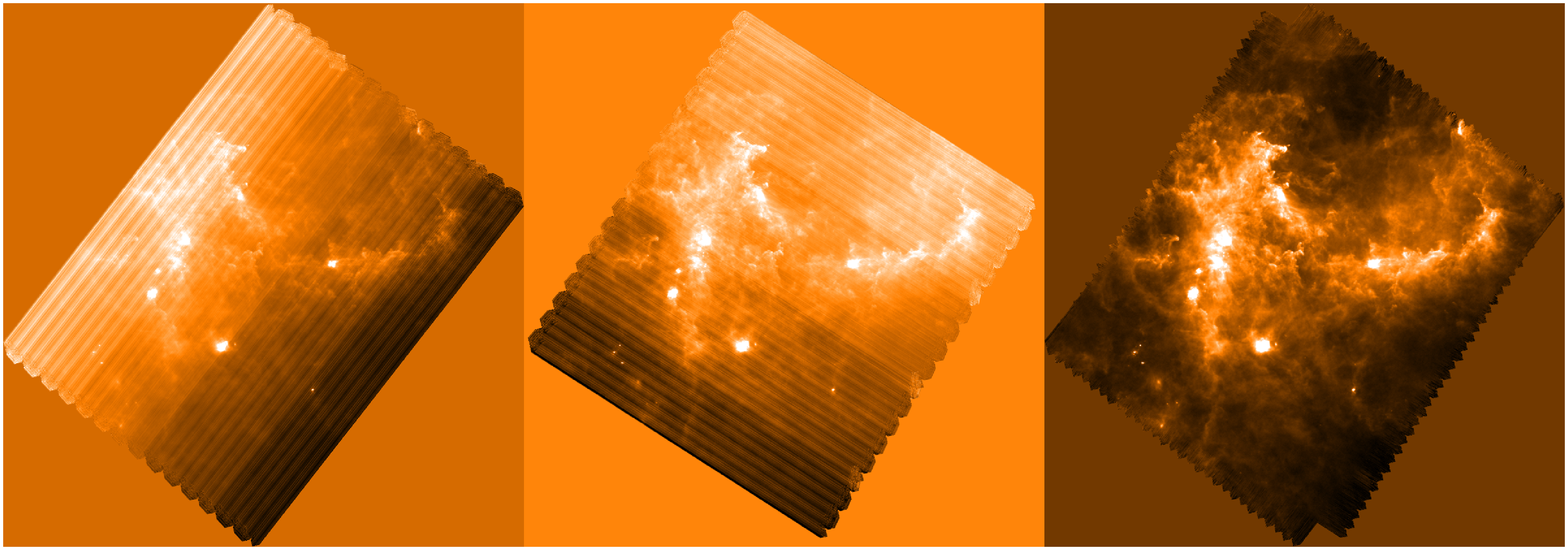}{f1_p043}{Maps of the scan (left) and cross-scan (center) observations of the Rosette cloud previous to any JScanam processing. Strong signal drifts are visible perpendicular to each scan direction. The right panel shows that the $1/f$ noise is removed after the JScanam automatic processing. Data from the HOBYS Key Programme (PI F.\@ Motte).}
     
In order to solve this problem, most of the PACS observations were done using a scan plus cross-scan observing strategy, where the region of interest is observed first in a given scan direction and later is scanned again in the perpendicular direction. In theory, this observing mode provides enough information and redundancy to separate the $1/f$ noise from the extended emission. 

\section*{JScanam}
Several map-making methods (JScanam, MadMap, Unimap, Scanamorphos, Sanepic) have been applied to reduce PACS photometer observations. Many of them can be run directly from the Herschel Interactive Processing Environment \citep[HIPE,][]{Ott2010}. Here we present some recent results from the JScanam map-maker, a HIPE/Java implementation of the Scanamorphos techniques initially developed and implemented in IDL by H{\'e}l{\`e}ne Roussel \citep{Roussel2013}. JScanam is currently the default mapper for reducing PACS photometer observations and its automatic pipeline results can be easily inspected and downloaded from the Herschel Science Archive\footnote{\url{http://www.cosmos.esa.int/web/herschel/science-archive}} and the ESA Sky interface \citep{Merin2015}.

The algorithm exploits the redundancy naturally provided by on-the-fly mapping with multi-detector arrays at at least two distinct scanning directions. Since each position on the sky is scanned by multiple bolometers at several different epochs, it is possible to disentangle the varying noise from the constant sky signal, with minimal assumptions about the noise.

\articlefigure{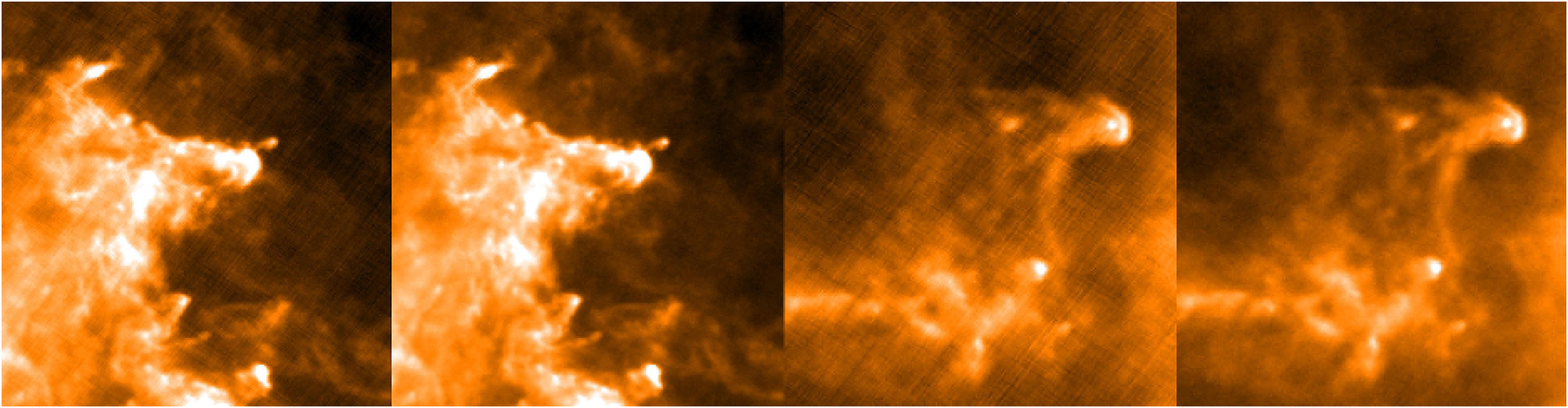}{f2_p043}{Detailed view of two regions in the Rosette cloud showing that the $1/f$ noise is also removed at the shorter spatial scales. The extended emission and the point sources remain unaffected. Data from the HOBYS Key Programme (PI F.\@ Motte).}

Although JScanam was initially based on Scanamorphos, during the last three years the development of both mappers has evolved independently (new pipeline steps, slightly different approaches to the same fundamental concepts). In particular, a strong effort has been put to make the JScanam automatic pipeline robust enough to deal with all kind of map sizes (from few square arcminutes to several square degrees) and astronomical environments (from Galactic star-forming clouds to cosmological fields), and to maximize the execution speed and memory efficiency.

\section*{Pipeline steps}

The JScanam processing pipeline starts from the scan and cross-scan calibrated Level~1 frames products. These are three-dimensional signal cubes, where the first two dimensions are used to indicate the row and column of the PACS photometer array pixels, and the third dimension is a temporal dimension \citep{Wieprecht2009}. At each time the array is at a different sky position, so this last dimension can also be seen as a spatial coordinate. 

The first pipeline steps are dedicated to identify and mask several effects that can decrease the final map quality: scan and cross-scan turnarounds where the scan velocity is not constant, strong signal jumps due to cosmic ray hits on the detectors, and possible exponential drifts introduced in the signal by the calibration blocks. At the same time a source mask is created. This mask is used for the rest of the pipeline to prevent cleaning artifacts produced by small scan and cross-scan spatial misalignments. The $1/f$ noise is then removed iteratively, starting from the longest spatial scales (higher noise power, longer timescales) and finishing at a scale of the order of the observation point spread function. After removing the $1/f$ noise, signal glitches are masked. Finally, the combination of the scan and cross-scan frames is projected on the sky to generate the clean map (the Level 2.5 product in Herschel/PACS terminology). 

Please, refer to the PACS Data Reduction Guide\footnote{\url{http://herschel.esac.esa.int/hcss-doc-13.0}} for a more in-depth description of the different pipeline steps and PACS product definitions. 

\articlefigure{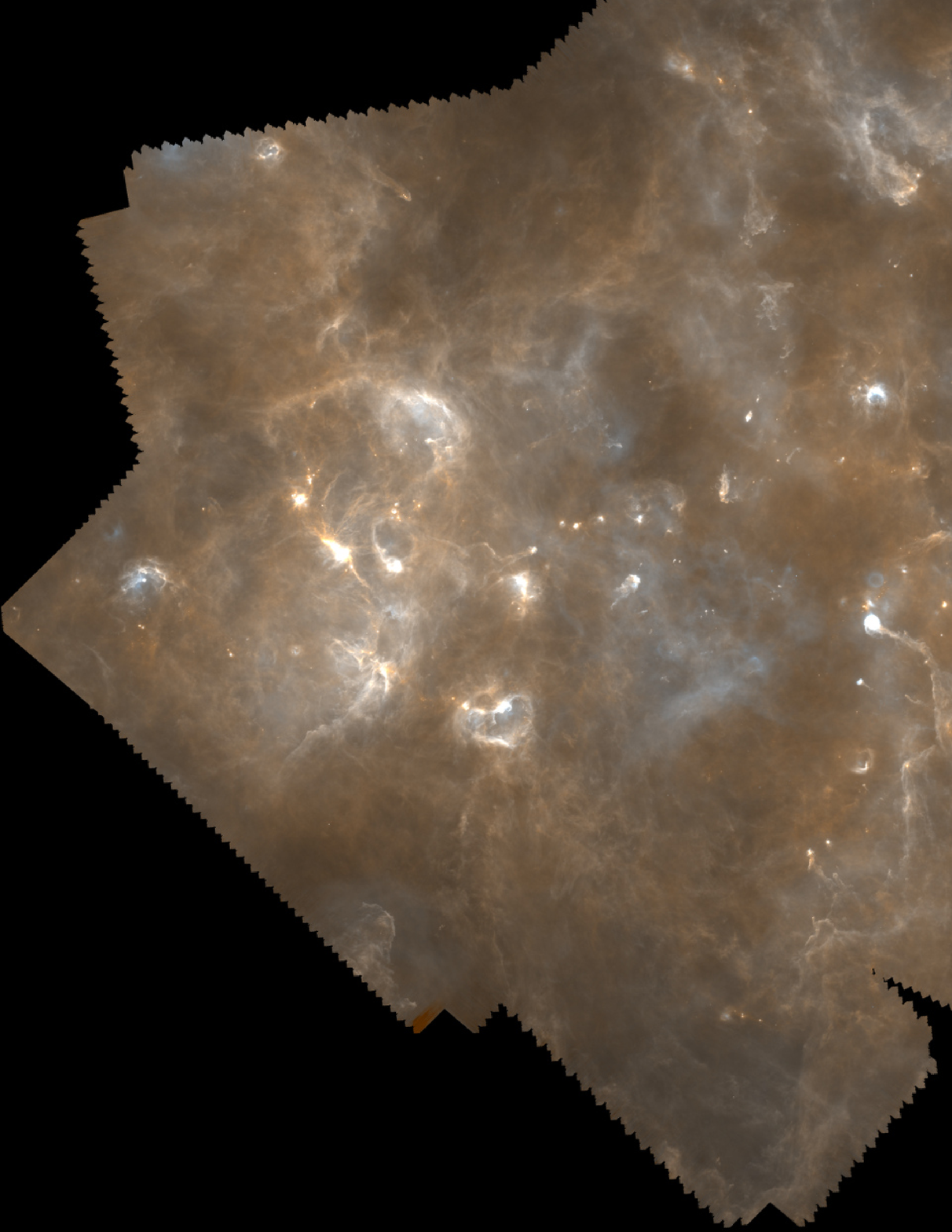}{f3_p043}{Composite map of the Cygnus-X massive star forming region (blue: 70$\mu$m, red: 160$\mu$m). The map, 8.0 degrees wide and 6.6 high, combines data from multiple programmes ({\tt OT2\_smolinar\_7}, {\tt KPGT\_fmotte\_1}, {\tt GT2\_pandre\_5} and {\tt OT2\_jhora\_2}). Each dataset has been independently reduced. As a final step, the regions were aligned (using only a constant offset) to have the same background signal.}

\section*{Current status}

The JScanam pipeline has entered a stable phase. Current development efforts are mostly going into the generation of PACS photometer Highly Processed Data Products. New methods to match and align the background continuum of different overlapping maps have been implemented and will be available in the next HIPE~14 release. Figure~\ref{f3_p043} shows an example result from the application of these methods to several PACS observations coming from 4 different programmes.

\bibliography{P043}

\end{document}